\pdfoutput=1
\documentclass{aastex}
\shorttitle{APO and CARMA Observations of Mol 121}
\begin{document}

\title{Massive Star Formation, Outflows, and Anomalous H$_2$ Emission in Mol 121 (IRAS 20188$+$3928)}

\author{Grace Wolf-Chase\altaffilmark{1} and Kim Arvidsson}
\affil{Astronomy Department, Adler Planetarium, 1300 S. Lake Shore Drive,
Chicago, IL 60605}
\email{gwolfchase@adlerplanetarium.org}

\author{Michael Smutko}
\affil{Center for Interdisciplinary Exploration and Research in Astrophysics (CIERA) \& Dept. of Physics \& Astronomy, Northwestern University, 2145 Sheridan Rd., Evanston, IL 60208}

\and

\author{Reid Sherman}
\affil{Dept. of Astronomy \& Astrophysics, University of Chicago, 5640 S. Ellis Ave.,
Chicago, IL 60637}

\altaffiltext{1}{Dept. of Astronomy \& Astrophysics, University of Chicago, 5640 S. Ellis Ave.,
Chicago, IL 60637}

\begin{abstract}

We have discovered 12 new molecular hydrogen emission-line objects (MHOs) in the vicinity of the candidate massive young stellar object  \object{Mol 121}, in addition to five that were previously known. H$_2$ 2.12-$\mu$m/H$_2$ 2.25-$\mu$m flux ratios indicate another region dominated by fluorescence from a photo-dissociation region (PDR), and one region that displays an anomalously low H$_2$ 2.12-$\mu$m/H$_2$ 2.25-$\mu$m flux ratio ($<$1) and coincides with a previously reported deeply embedded source (DES). Continuum observations at 3 mm reveal five dense cores; the brightest core is coincident with the DES. The next brightest cores are both associated with cm continuum emission. One of these is coincident with the {\it IRAS} source; the other lies at the centroid of a compact outflow defined by bipolar MHOs. The brighter of these bipolar MHOs exhibits [\ion{Fe}{2}] emission and both MHOs are associated with CH$_3$OH maser emission observed at 95 GHz and 44 GHz. Masses and column densities of all five cores are consistent with theoretical predictions for massive star formation. Although it is impossible to associate all MHOs with driving sources in this region, it is evident that there are several outflows along different position angles, and some unambiguous associations can be made. We discuss implications of observed H$_2$ 2.12-$\micron$/H$_2$ 2.25-$\micron$ and [\ion{Fe}{2}] 1.64-$\micron$/H$_2$ 2.12-$\micron$ flux ratios and compare the estimated total H$_2$ luminosity with the bolometric luminosity of the region. We conclude that the outflows are driven by massive young stellar objects embedded in cores that are likely to be in different evolutionary stages.

\end{abstract}


\section{INTRODUCTION}

Outflows play a critical role in the formation of isolated low-mass stars throughout the embedded accretion phases of evolution. They are thought to be driven by magnetic stresses in accretion disks that eject some of the inflowing matter and carry off angular momentum (e.g., Shu et al. 1994; Pudritz et al. 2006; Matsakos et al. 2009). In contrast, the role of outflows in the formation of stars $>$ 8 M$_{\odot}$ remains elusive, because these stars are rare, form in tight clusters along with low- and intermediate-mass stars that may be in different stages of evolution, lie in heavily obscured regions often at distances greater than several kiloparsecs, and may be evolving on dynamical time scales less than a few thousand years (e.g., Zinnecker \& Yorke 2007). Large surveys with single-antenna microwave telescopes indicate that  energetic molecular outflows are present in massive star-forming regions (e.g., Zhang et al. 2005), but it is not generally possible to link individual outflows with driving sources, even with the current generation of millimeter-wave interferometers. 

The situation is somewhat better in the near-infrared, where
the H$_2$ 1-0 S(1) line at 2.12 $\micron$ is a particularly powerful tracer of shocks in molecular outflows (Davis et al. 2010 and references therein; Smith 2012). The current generation of sensitive wide-field near infrared imagers enables large surveys of massive star-forming regions to be conducted at arcsecond or sub-arcsecond resolution in reasonable periods of observing time. Using the  Apache Point Observatory (APO) 3.5-m telescope, Wolf-Chase et al. (2012, in preparation) have conducted a narrow-band near-infrared survey for shocked emission in 28 massive star-forming regions that exhibit energetic molecular outflows (Zhang et al. 2005). In Wolf-Chase et al. (2012, hereafter WSS12), we presented near-infrared results for Mol 160 (IRAS 23385+6053), along with 3-mm continuum and line observations obtained with the Combined Array for Research in Millimeter-wave Astronomy (CARMA). In this paper, we present narrowband, $\sim$ 1$^{\prime\prime}$-resolution near-infrared APO images and $\sim$ 2$^{\prime\prime}$-resolution CARMA 3-mm continuum and line observations for Mol 121 ({\it IRAS} 20188+3928).

Mol 121 has a spectral energy distribution (SED) resembling an ultra-compact (UC) \ion{H}{2} region and is associated with radio continuum emission (Wood \& Churchwell 1989; Molinari et al. 1998). Varricatt et al. (2010) included it in a NIR imaging survey of 50 regions containing intermediate- and high-mass YSO outflow candidates. They present an overview of background observations of this object in the appendix to their paper (A37) and note that prior distance estimates to Mol 121 span an order of magnitude (0.3 - 3.91 kpc). Mol 121 (MSX6C designation G077.4622+01.7600) is categorized in the Red MSX Survey (RMS) as Class \ion{H}{2}/YSO, and is among the RMS objects that have uniquely constrained kinematic distances and good SED fits (Mottram et al. 2011, Table 1). In this paper, we assume the RMS catalog distance of 1.7 kpc and bolometric  luminosity of 8.2$\times 10^3$ L$_{\odot}$. 

\section{OBSERVATIONS AND DATA REDUCTION}

\subsection{Apache Point Observatory: NICFPS}

We obtained narrowband, near-infrared images of Mol 121 using the Near-Infrared
Camera and Fabry-Perot Spectrometer (NICFPS) on the Astrophysical
Research Consortium (ARC) 3.5-m telescope at the APO in Sunspot, NM on UT 2008 Oct 13. 
We used the following filters and total integration times:  H$_2$ 2.12-$\micron$ (1740 s), H$_2$ 2.25-$\micron$ (1740 s), and [\ion{Fe}{2}] 1.64-$\micron$ (1800 s).  We also used two narrowband filters to allow for continuum subtraction in the final images: an H$_2$-continuum filter centered on 2.13 $\micron$ (1740 s) and an [\ion{Fe}{2}]-continuum filter centered on 1.65 $\micron$ (1800 s). Our basic data acquisition, reduction, and calibration methods are described in WSS12. 

Thirteen isolated stars having the best 2MASS photometry flags were chosen in the narrow-band line images as well as in the corresponding narrow-band continuum images.  We performed aperture photometry on these stars using the DAOPHOT package in IRAF with an aperture radius of 12 pixels, $\sim$4 times the FWHM, since the seeing was $\sim$1$^{\prime\prime}$ in each image. The continuum images were scaled using a scaling factor and then subtracted from the narrow-band line images so that on average the stars disappear. The scaling factor was chosen so that the fractional residual fluxes of the stars within the aperture are as close to zero as possible in the resulting continuum-subtracted image. This procedure worked well except for the brightest stars in the field, which show small residual artifacts. We flux calibrated our data using equations (1) \& (2) from WSS12 with  $d\lambda = 5.5$ nm for the [\ion{Fe}{2}] 1.64-$\micron$ filter (Apache Point Observatory 2010).

Since the H$_{2}$ sources can have irregular morphologies, we performed irregular aperture photometry using the highly flexible kang\footnote{http://www.bu.edu/iar/kang/} software version 1.3. The apertures were defined using polygons. For each aperture in each image, we drew ten background regions. These background regions were chosen so they sample the background as close to the aperture as possible, and were roughly of the same size as the apertures. We then used the average background-subtracted flux within each aperture for subsequent calculations. The uncertainties in Tables 1 \& 2 correspond to $\pm 1 \sigma$, and are the combinations of three categories of uncertainty; calibration, subtraction, and aperture. The calibration uncertainty incorporates the 2MASS catalog K$_{s}$ magnitude uncertainties, our instrumental stellar magnitude uncertainties and the standard deviation for the zero-point correction. The subtraction uncertainty is estimated using the standard deviation from zero for the fractional residual fluxes in the continuum subtraction procedure. The aperture uncertainty is the combination of photon noise within the aperture and background regions and the standard deviation from the ten background regions used for each aperture.

\subsection{CARMA}

We observed a field around Mol 121 approximately 1.5$^{\prime}$ across with CARMA on UT 2010 March 11. The interferometer was in the C configuration with baselines ranging from 26-370 meters. We simultaneously observed the continuum and the CH$_3$OH J=8$\rightarrow$7 line at 95.169 GHz. 
Two spectral bands were configured for maximum continuum bandwidth, covering 468 MHz each, for a total of 1.872 GHz (including both sidebands), on either side of a central frequency of 93.585 GHz. The band observing CH$_3$OH was in the upper side band with a 31-MHz bandwidth and 63 channels, resulting in a velocity resolution of 1.538 km s$^{-1}$. The passband calibrator was 1927$+$730, the flux calibrator was MWC349, and the phase calibrator was 2007$+$404. The total integration time on source was 4.92 hours. The size (FWHM) of the synthesized beam was 1.967$^{\prime\prime}\times$1.516$^{\prime\prime}$. We reduced and calibrated the data as described in WSS12. 

\section
{RESULTS}

\subsection{H$_2$ \& [\ion{Fe}{2}] Emission}

Figures 1-3 present our narrow-band NIR images. Our H$_2$ 2.12-$\micron$, H$_2$ 2.25-$\micron$, [\ion{Fe}{2}], and continuum-subtracted line  images are shown in Figure 1. Figure 2 displays the continuum-subtracted H$_2$ 2.12-$\micron$ image with the regions used to identify molecular hydrogen emission-line objects (MHOs), compact emission-line sources thought to be associated with outflows. Two other emission features that are not MHOs are also identified:  fluorescent gas in a photo-dissociation region (PDR) and a region where the H$_2$ 2.12 $\micron$/2.25 $\micron$ flux ratio is less than one (marked as `Anomaly'). The anomalous H$_2$ emission is discussed further in \S 4.1. For the MHOs, we follow the numbering scheme of the on-line catalog of these objects, hosted by the Joint Astronomy Centre in Hawaii  (Davis et al. 2010). In all, we identify 12 new MHOs in addition to five that have been catalogued previously (see Figure 1: bottom left panel, Figure 2, \& Table 1). 
The feature originally identified as MHO 867 by Varricatt et al. (2010) actually encompasses both the PDR and the bow-shaped feature we associate with the MHO. Although the {\it IRAS} source appears to be offset at the eastern edge of the PDR in the H$_2$ 2.12-$\micron$ and [\ion{Fe}{2}] images, it is centrally located in the H$_2$ 2.25-$\micron$ images (Figures 1: top and bottom middle panels) and the 3-color Wide-field Infrared Survey Explorer (WISE: Wright et al. 2010) image presented in Figure 4. The presence of shocked gas to the west of the {\it IRAS} source may explain the enhancement of H$_2$ 2.12-$\micron$ and [\ion{Fe}{2}] emission in this direction.

Table 1 lists the MHO designations (column 1), areas used in the aperture photometry (column 2), positions of the peak emission (columns 3 \& 4), H$_2$ 2.12-$\micron$ line fluxes (column 5), H$_2$ 2.25-$\micron$ line fluxes (column 6), and H$_2$ 2.12-$\micron$/H$_2$ 2.25-$\micron$ flux ratios (column 7).  Columns 8$-$10 present results for fluxes integrated over 2-pixel circular apertures centered on 2.25-$\micron$ peaks. Where the H$_2$ 2.25-$\micron$ line flux is below 3 times the noise (3$\sigma$) within an aperture, the H$_2$ 2.25-$\micron$ line is regarded as undetected and no value is given in columns 6 or 9. In these cases, lower limits to the H$_2$ 2.12-$\micron$/H$_2$ 2.25-$\micron$ flux ratios are calculated using the 3$\sigma$ value for that aperture as an upper limit for the H$_2$ 2.25-$\micron$ line flux. The difference in extinction at 2.12 $\micron$ and 2.25 $\micron$ is very small, only a tenth of a magnitude (assuming A$_{\lambda} \sim {\lambda}^{-1.7}$), yielding a corresponding brightness ratio of $\sim$1.1, which has a negligible effect on the flux ratios. 

The expected H$_2$ 2.12-$\micron$/2.25-$\micron$ line ratio for UV excitation (as in PDRs) is $\sim$ 1.9. It is significantly higher for H$_2$ excited by shocks, with a ratio of $\sim$ 3 - 20 for different types of shocks (e.g., Black \& van Dishoeck 1987; Smith 1994; Gredel \& Dalgarno 1995; Smith 1995; Smith et al. 2003). Two types of shocks are observed in outflows from star-forming regions. Continuous (or C-type) shocks occur when the shock is magnetically cushioned by a relatively high transverse magnetic field, such that the shock thickness is broad and deceleration of gas is relatively gradual, resulting in lower excitation and temperatures than in Jump (or J-type) shocks. In the latter, the gas undergoes a sharp increase in temperature within a narrow zone, where molecules are dissociated (e.g., Barsony et al. 2010; Smith 2012).

The observed line ratios for objects listed as MHOs are consistent with shock excitation, but not for the PDR and Anomaly. Since the high flux ratios for many of the MHOs may reflect the fact that 2.25-$\micron$ emission is not detected over the full extent of the apertures used for the 2.12-$\micron$ emission (see Table 1, columns 5-7), we also computed flux ratios using a circular aperture of radius 2 pixels centered on the  2.25-$\micron$ emission peaks (see Table 1, columns 8-10). A comparison of columns 10 and 7 indicates that although H$_2$ 2.12-$\micron$/2.25-$\micron$ flux ratios are typically lower in the 2-pixel radius apertures, the ratios are still consistent with shock excitation. 

The  [\ion{Fe}{2}] 1.64-$\micron$ line probes fast, dissociative shocks with velocities $>$ 50 km s$^{-1}$. When it is observed in jets from low- and intermediate-luminosity YSOs, it is typically localized in compact spots, representing only a small fraction of the region emitting in H$_2$, and its presence does not appear to correlate strictly with the evolutionary stage of the driving source (Caratti o Garatti et al. 2006). Fewer [\ion{Fe}{2}]  observations have been reported for high-luminosity objects, in part due to the greater extinction at 1.64 $\micron$. Curiously, Caratti o Garatti et al. (2008) found that the only [\ion{Fe}{2}] emission toward the H$_2$ jet associated with the Massive Young Stellar Object (MYSO) {\it IRAS} 20126$+$4104 is close to the source. Table 2 lists the three [\ion{Fe}{2}] emission-line features we detected. The first two features are associated with MHO 864, and the third feature is associated with, and largely overlaps, MHO 867, the bright bow- or `bean'-shaped feature that opens toward the {\it IRAS} source. In MHO 864, the two [\ion{Fe}{2}] emission-line features lie along the inferred jet axis of the outflow.  One is located at the apex of the MHO and the other is close to core D, which presumably harbors the driving source of the outflow (see Figure 2 and \S 4.2.4). In MHO 867, the morphology of the [\ion{Fe}{2}] emission is similar to the MHO, but over a smaller region. To investigate excitation conditions in these MHOs, we computed line ratios in 2-pixel radius apertures along the major and minor axes of MHO 864, and along and transverse to the bow-shaped structure of MHO 867. Placement of these apertures is indicated in Figure 3 and emission line flux ratios are presented in Table 3. For the apertures with H$_2$ 2.25-$\micron$ detections, 2.12 $\micron$/2.25 $\micron$ flux ratios range from 6.8 - 14.5. 1.64-$\micron$/2.12-$\micron$ flux ratios range from 0.03 - 0.9 for MHO 864 apertures and 0.5 - 1.9 for MHO 867 apertures. We interpret these results in \S 4.1.

The terms ``Green Fuzzies''  and ``Extended Green Objects'' (EGOs) have been applied to regions of extended 4.5-$\micron$ emission that are often associated with shocked H$_2$ emission from outflows (e.g., Cyganowski et al. 2008). The nomenclature reflects the use of green as the representative color for the InfraRed Array Camera (IRAC: Fazio et al. 2004) 4.5-$\micron$ band in Spitzer Space Telescope 3-color images produced from Galactic Legacy Infrared Mid-Plane Survey Extraordinaire observations (GLIMPSE: Benjamin et al. 2003; Churchwell et al. 2009). Figure 4 uses a similar 3-color scheme to represent the WISE 3.4, 4.6, \& 12-$\micron$ bands. Here, objects that are bright in the 4.6-$\micron$ band appear to be green. MHOs 864, 865, \& 866 are the only MHOs that are clearly prominent in the 4.6-$\micron$ band. MHOs 864 \& 865 have the highest H$_2$ 2.12-$\micron$ fluxes in Mol 121. On the other hand, there are several patches of faint 4.6-$\micron$ emission that do not correspond to MHOs. We note that this result is similar to that of Lee et al. (2012), who imaged 34 Spitzer GLIMPSE EGOs with the United Kingdom Infrared Telescope (UKIRT) and noted that shocked molecular hydrogen emission is typically more extended than EGOs and that EGOs occasionally trace scattered continuum light (see also Barsony et al. 2010). This suggests that although the 4.5-$\micron$ \& 4.6-$\micron$ bands may be useful in identifying features associated with bright outflows, they are not always dominated by shocked outflow emission.

\subsection{3-mm Continuum Emission}

Our CARMA 3-mm continuum data reveal five cores in the Mol 121 region. Figure 5 shows contours for the 3-mm continuum emission overlaid on the relevant portion of our continuum-subtracted H$_2$ 2.12-$\micron$ image. Table 4 lists physical properties of the cores. Assuming a distance of 1.7 kpc, deconvolved core diameters, $(D_{max}D_{min})^{1\over 2}$, range from 790 AU - 3500 AU. The cores are discussed individually in \S 4.2. 

\subsubsection{Estimating Core Mass \& Column Density}

Assuming the dust emission is optically thin at 3 mm, the mass of a core is calculated in the following manner:

$$ M = {d^2S_\nu R \over B_\nu(T_D)\kappa_\nu} \eqno{(1)} $$ 

Here, $d$ is the distance to the source, $S_{\nu}$ is the total flux density, $R$ is the gas-to-dust ratio, $B_{\nu}(T_D)$ is the Planck function at dust temperature $T_D$, and $\kappa_\nu$ is the dust opacity per gram of gas (e.g., Enoch et al. 2007). We use $\kappa_\nu$ = 0.178 cm$^2$ g$^{-1}$, the value obtained by extrapolating from Table 1 in Ossenkopf \& Henning (1994) using a power law ($\kappa \propto \lambda^{-\beta}$) with $\beta$ = 1.8, and assuming dust grains with thin ice mantles coagulated for 10$^5$ yr at a gas density of 10$^6$ cm$^{-3}$. Typically, dust temperatures in protostellar cores are assumed to be $\sim$ 20 K (e.g., Rosolowsky et al. 2010; Lee et al. 2011; Chen et al. 2012), but dust temperatures in PDRs and some massive protostellar cores are closer to 30 K (e.g., Molinari et al. 2008; Anderson et al. 2012), and Enoch et al. (2007) note that plausible values might vary from 5$-$30 K.  The SED of {\it IRAS} 20188$+$3928 has been fit with a cold dust component of 39 K and warm dust component of 100 K (McCutcheon et al. 1995). Most of the cores detected by CARMA are probably colder than this, so we have opted to calculated masses for T$_D$ = 20 K \& 40 K (Table 3, columns 7 \& 8). For T$_D$ = 40 K, derived masses are $\sim$ 2.1 times lower than for T$_D$ = 20 K. 

We calculate the column density from the peak flux density $I_\nu$ in a similar manner to Enoch et al. (2007):

$$ N(H_2)={I_\nu R \over {\Omega_{mb} m_{H_2} \kappa_\nu B_\nu(T_D)}} \eqno{(2)} $$

Here, $R$, $\kappa_\nu$, and  $B_{\nu}(T_D)$, are the same as in equation (1), $\Omega_{mb}$ is the size of CARMA's main beam at 3 mm, and $m_{H_2}$ is the mass of molecular hydrogen. Columns 9 \& 10 in Table 3 show the results of this calculation for both dust temperatures used in the mass calculation.

Theoretical work predicts a mass column density threshold $\ge$ 1 g cm$^{-2}$ for massive star formation (Krumholz \& McKee 2008; Krumholz et al. 2010). The mass column density is the column density $N(H_2)$ multiplied by $m_{H_2}$ and a factor of 1.36 to account for elements heavier than hydrogen (Simon et al. 2001). Regions where intermediate mass stars are thought to be forming typically have mass column densities $\sim$ 0.1 - 0.5 g cm$^{-2}$ (e.g., Arvidsson et al. 2010; Wolf-Chase et al. 2003; Wolf-Chase, Walker, \& Lada 1995). All of the CARMA cores have mass column densities significantly $>$ 1 g cm$^{-2}$, the smallest value being $\sim$ 6 g cm$^{-2}$ for core E at T$_D$ = 40 K. 

\subsection{CH$_3$OH Masers}

We have discovered three 95-GHz CH$_3$OH masers in the Mol 121 region, coincident with MHOs 864, 865, and 892. They are all unresolved by CARMA's 2$^{\prime\prime}$ resolution, and each appear in two contiguous velocity channels. There are very slight ($<$1$^{\prime\prime}$) shifts in position of the masers between channels, and that, combined with our relatively coarse 1.538 km s$^{-1}$ channel width, means that our masers may each be multiple, tightly clustered masers, but without better spatial and spectral resolution we will set aside this possibility. For each maser, we measured the total flux and estimated the central velocity, and these results are presented in Table 5.

One 44-GHz Class I CH$_3$OH maser was previously discovered in the Mol 121 region (S. Kurtz, private communication), at the same location as the brightest 95-GHz maser and with a velocity of 2.3 km s$^{-1}$. The channel width of the 44-GHz data was 0.166 km s$^{-1}$, accounting for the discrepancy with our estimated velocity. The 44-GHz entries in Table 5 were identified from VLA data provided by Kurtz and are discussed in \S 4.4.

\section{DISCUSSION}

\subsection{H$_2$ \& [\ion{Fe}{2}] Excitation}

Lorenzetti et al. (2002) found that [\ion{Fe}{2}] 1.64-$\micron$/H$_2$ 2.12-$\micron$ flux ratios for knots associated with a jet from IRS 8 in the Vela Molecular Ridge ranged from $\sim$ 0.1 - 1.1 (uncorrected for extinction). Since the [\ion{Fe}{2}] and H$_2$ emitting regions are unsurprisingly physically distinct, they computed flux ratios based on two sets of zones, defined by the [\ion{Fe}{2}] peak emission and the H$_2$ peak emission. They argue that these high ratios require the presence of a fast J-shock component, which produces [\ion{Fe}{2}] 1.64-$\micron$/H$_2$ 2.12-$\micron$ flux ratios in the range 0.5 - 2 for a wide range of parameters (Hollenbach \& McKee 1989). Given that the effects of extinction make our observed [\ion{Fe}{2}] 1.64-$\micron$/H$_2$ 2.12-$\micron$ flux ratios lower limits, this suggests the presence of J-shocks in both MHOs 867 \& 864; however, the high observed H$_2$  2.12-$\micron$/2.25-$\micron$ ratios across all of the apertures with 2.25-$\micron$ detections (6.8 - 14.5) in Table 3 suggest the presence of C-shocks as well. For the high excitation temperatures associated with J-shocks, one expects H$_2$  2.12-$\micron$/2.25-$\micron$ $\sim$ 3 $-$ 4, and for the lower-excitation C-shocks, $\sim$ 6 $-$ 20. Above 20, the excitation temperature is $<$ 1000 K, and observable emission is not expected (Smith 1994, 1995; Smith et al. 2003). In fact, both types of shocks may be found in jets from YSOs. Typically, the molecular emission arises from the flanks of a bow shock, and the atomic or ionized component originates close to the apex (Smith 2012).

Neither fluorescence nor shocks can explain a H$_2$ 2.12-$\micron$/2.25-$\micron$ flux ratio $<$ 1, which is the case for the ``Anomaly'' listed in Table 1 and indicated on Figure 2. We hand-checked each image for possible saturation effects near the bright {\it IRAS} source. Our combination of narrow-band filters and short exposure times (30 s) kept even the brightest sources below 30k ADU, well within the linear range of the detector (3\% linear to 54.4k ADU) and far below its full-well capacity of more than 100k ADU. Therefore, we conclude that this effect is real. To the best of our knowledge, this is the first time an observed  H$_2$ 2.12-$\micron$/2.25-$\micron$ flux ratio $<$ 1 has been reported. Although high-resolution NIR spectroscopy is essential to determining the nature of this emission, we note that Dove et al. (1987) found that certain circumstances could produce population inversions in para-hydrogen transitions. They suggested future work to include the ortho-hydrogen transitions and more completely model the processes occurring in shock-heated clouds. Another intriguing possibility is that the anomalous emission comes from the immediate front of a fast dissociative shock (T$\ge$10,000 K), where statistical equilibrium is not reached (M. Smith, private communication). If statistical equilibrium is not achieved, the 2.12-$\micron$/2.25-$\micron$ ratio should be the ratio of the transition probabilities. The transition probabilities for the H$_2$ 2.12-$\micron$ 1-0 S(1) and H$_2$ 2.25-$\micron$ 2-1 S(1) transitions are 3.47$\times 10^7$ s$^{-1}$ and 4.98$\times 10^7$ s$^{-1}$, respectively (Turner et al. 1977). In this case, the expected ratio would be about 0.7. Although this is still somewhat larger than we calculate in Table 1, the ratio is affected by choice of aperture size. The Anomaly is bright in both line and continuum NIR  images (see Fig. 1), and it is coincident with a deeply embedded infrared source (DES)  identified from polarimetric patterns in the scattered NIR emission (Yao et al. 2000). The DES is also coincident with core C, the brightest 3-mm core in the Mol 121 region (see \S4.2.3).

\subsection{Dense Cores \& Outflows}

In this section, we discuss each of the five cores we detected with CARMA, previous related observations, and the possible association of cores with outflows, individually. There is direct evidence of massive star formation occurring in three of the five cores (A, C, \& D).  The remaining two cores (B \& E) may be starless. In the absence of significant external heating, these cores are likely to have temperatures $\le$ 20 K, in which case the derived masses and mass column densities suggest that they may be sites of future massive star formation. 
 
\subsubsection{Core A}

Core A is associated with the {\it IRAS} source and several signposts of massive star formation, including a compact \ion{H}{2} region observed at 3.6 cm and 6 cm (Jenness et al. 1995; Molinari et al. 1998), a PDR detected in our H$_2$ 2.12-$\micron$, 2.25-$\micron$, and [\ion{Fe}{2}] 1.64-$\micron$ observations (Figs. 1 \& 2) and evident in the WISE image (Fig. 4), and H$_2$O maser emission (Palla et al. 1991; Brand et al. 1994; Anglada et al. 1997). Furthermore, core A lies at or near the continuum peak identified through previous sub-millimeter and millimeter observations (FIR 2 - Jenness et al. 1995; McCutcheon et al. 1995; Curran \& Chrysostomou 2007; Motte et al. 2007).  Core A  may be associated with a $\sim$E-W outflow defined by several MHOs. In particular, the bow-shaped MHO 867 opens toward this source, as does MHO 938 further to the west (see Figures 2 \& 5). MHO 867 also exhibits strong [\ion{Fe}{2}] emission. There are several MHO groupings toward the east, which do not align in any obvious way (MHOs 939$-$942).

\subsubsection{Core B}

Core B has not been detected previously with millimeter or sub-millimeter observations, nor is it associated with any obvious signposts of star formation such as masers, radio continuum, or obvious outflow activity. It is located at the northern end of the bright PDR, coincident with an apparent dip in the NIR emission, which suggests high extinction towards this core. It may be prestellar in nature. If this is true, its temperature may be lower than we assumed, and thus its mass may be larger. 

\subsubsection{Core C}

Core C has not been detected previously with millimeter or sub-millimeter continuum observations, even though it has the largest integrated flux of the cores we detected with CARMA. It does, however, coincide with the location of the DES reported by Yao et al. (2000), as well the peak of the anomalous H$_2$ emission. Although Motte et al. (2007) mapped the entire Cygnus X molecular cloud complex in 1.2-mm continuum at IRAM, they detected only one dense core with their 11$^{\prime\prime}$ beam within the region we identify as a PDR associated with {\it IRAS} 20188$+$3928. Our cores A, B, \& C all lie within this region. Little et al. (1988) first detected a bipolar molecular outflow in the vicinity of {\it IRAS} 20188$+$3928 from CO and HCO$^+$ observations. The molecular outflow is aligned roughly N-S, with the redshifted gas to the north. Using a distance of 4 kpc, they noted the outflow properties were similar to other massive outflows such as Cep A and NGC 2071. Zhang et al. (2005) identified an outflow with the same orientation from CO 2$\rightarrow$1 observations at a similar resolution to the observations of Little et al. (1988). They derived significantly smaller outflow mass and energetic parameters; however, they used a distance of only 0.3 kpc, which is much smaller than the most recent estimate of the distance (Mottram et al. 2011). Zhang et al. (2005) noted that the outflow appears to be centered $\sim$6$^{\prime\prime}$ to the north of the {\it IRAS} source, coincident with the DES identified by Yao et al. (2000); however, given their $\sim$30$^{\prime\prime}$ beam and spatial sampling, it is impossible to identify the driving source(s). The K$_s$ image of Yao et al. (2000) shows NIR reflection nebulosity oriented generally N-S in the vicinity of the {\it IRAS} source; however, the polarization vectors  exhibit a different pattern in a NW-SE direction about the DES (coincident with the bar-shaped extension of the Anomaly evident in Figs. 2 \& 5). Yao et al. (2000) suggest this may indicate the direction of a newly-burst outflow from the DES.  The association of core C with the DES identified by Yao et al. (2000), strong millimeter flux, and corresponding lack of cm-continuum emission, make this object an excellent candidate MYSO in a pre-UC \ion{H}{2} phase of evolution. If this is the case, anomalous H$_2$ emission might be a signpost for the early outflow phase of a MYSO, but clearly high-resolution NIR spectroscopy is necessary to search for any such evolutionary trends. 

\subsubsection{Core D}

Core D is associated with a H$_2$O maser and faint continuum emission at 3.6 cm and 6 cm, indicative of the presence of a UC \ion{H}{2} region (Jenness et al. 1995; Molinari et al. 1998). Using the UKT14 receiver at the JCMT, Jenness et al. (1995) detected a submillimeter source (FIR 1)  at a position between our cores D \& E  from 450 $\micron$ \& 800 $\micron$ $\sim$ 18$^{\prime\prime}$-resolution observations. MHOs 864 \& 865, the brightest MHOs in the region (with the exception of MHO 867, which may be contaminated with fluorescent emission from the PDR), exhibit bipolar morphology in a NE-SW direction about core D. Both MHOs are associated with CH$_3$OH maser emission at 95 GHz \& 44 GHz. We have also detected [\ion{Fe}{2}] emission at two positions within MHO 864. Several MHOs lie near or along a slightly different position angle from the direction defined by MHOs 864 \& 865: to the NE, MHOs 863, 891 \& 892, and to the SW, MHO 937. MHO 892 is also coincident with CH$_3$OH maser emission at 95 GHz \& 44 GHz. Given the presence of both radio continuum emission and one or more well-collimated H$_2$ outflow, this object may be close to transitioning to a stage where an expanding UC \ion{H}{2} region destroys the MYSO(s)' accretion disk. The bright MHO 866 lies directly to the NW of core D, but its association is unclear. 

\subsubsection{Core E}

The position of core E coincides with the position of a dense core detected by Motte et al. (2007) in 11$^{\prime\prime}$-resolution 1.2-mm continuum observations using the IRAM 30-m telescope. For an assumed distance of 1.7 kpc and temperature of 20 K, Motte et al. (2007) derive a core mass of 40 M$_{\odot}$ and density, n(H$_2$), of 3.5$\times 10^5$ cm$^{-3}$. For the same temperature, we derive a core mass of 12.8 M$_{\odot}$. The difference can be explained if the CARMA observations resolve out some of the large-scale structure revealed in sub-millimeter JCMT maps (McCutcheon et al. 1995; Jenness et al. 1995; Curran \& Chrysostomou 2007). MHOs 893$-$897 lie in the vicinity of  core E, but it is not possible to associate any of these with a driving source presently. 

\subsection{H$_2$ Luminosity}

WSS12 noted that the MYSO, Mol 160, exhibits a very high L$_{H_2}$/L$_{bol}$ ratio, falling approximately a factor of two above the relationship that was determined for low-mass YSOs (Caratti o Garatti et al. 2006) and extended to MYSOs based on results for IRAS 20126$+$4104 (Caratti o Garatti et al. 2008). Although it is not generally possible to link individual outflows with driving sources in massive star-forming regions, and caution must be exercised in attempting to draw conclusions about the nature of MYSO outflows from the L$_{H_2}$/L$_{bol}$ relationship alone (WSS12), it is instructive to compare the total H$_2$ luminosity from all detected outflows to the total bolometric luminosity of the Mol 121 region. If the H$_2$ flows are produced by low- or intermediate-mass YSOs, one would expect the H$_2$ emission to be under-luminous with respect to the bolometric luminosity, which comes primarily from the {\it IRAS} source. 

There are many uncertainties in estimating the total H$_2$ luminosity, including the distance to Mol 121/{\it IRAS} 20188$+$3928, the extinction toward individual H$_2$ knots, and the ratio of total  H$_2$ to H$_{2.12}$ luminosity. Nevertheless it is possible to place reasonable limits on the assumptions. Without spectra of the MHOs, we can only estimate the effects of extinction. Although the extinction is very large in the dense cores, most of the MHOs are located in the more extended cloud. It seems reasonable to assume the column density derived for the parsec-scale NH$_3$ clump in this region by Anglada et al. (1997), N(H$_2$) = 5.5$\times 10^{22}$ cm$^{-2}$. Using the relation A$_V$/N(H$_2$) $\approx$ 5.3 $\times 10^{-22}$ cm$^2$ mag yields an estimate of the visual extinction A$_V$ $\approx$ 29 mag. Further assuming an optical parameter for dense clouds of R$_V$ = 5, yields A$_{2.12}$ $\approx$ 3.6 mag (Allen 2000).

The luminosity of the H$_2$ 2.12 $\micron$ emission is given by:

$$ L_{2.12} = 4\pi d^2 F_{2.12} \times 10^{0.4A_{2.12}} \eqno{(3)} $$

F$_{2.12}$ = 4.59 $\times$ 10$^{-16}$ W m$^{-2}$, summed over all the MHOs listed in Table 1 (including the PDR and Anomaly increases F$_{2.12}$ by a factor of $\sim$ 2.5.) Depending upon the temperature of the shocked H$_2$, the 2.12-$\micron$ luminosity is typically $\sim$ 5-10\% of the total ro-vibrational H$_2$ emission, L$_{H_2}$ (e.g., Caratti o Garatti 2006, 2008). This yields L$_{H_2}$ $\sim$ 11.7 - 23.3 L$_{\odot}$ for the combined MHO luminosity. The bolometric luminosity of {\it IRAS} 20188$+$3928, calculated from the far-infrared SED and assuming d = 1.7 kpc, is  8.2$\times$10$^3$ L$_{\odot}$ (Mottram et al. 2011). These results place Mol 121 a factor of 1.5 to three above the L$_{H_2}$/L$_{bol}$ relationship determined by Caratti o Garatti et al. (2006, 2008). This is very similar to the result obtained by WSS12 for Mol 160, and strongly suggests that the H$_2$ flows are driven by massive YSOs. 

\subsection{CH$_3$OH Masers}

Class I CH$_3$OH masers are well-correlated with molecular outflows in massive star-forming regions (Cyganowski et al. 2009 and references therein; Fontani et al. 2010) and with the presence of compact 3-mm continuum emission (Schnee \& Carpenter 2009), suggesting that these masers are signposts of an early stage in the evolution of a MYSO before an expanding UC \ion{H}{2} region has destroyed the accretion disk.

The strongest CH$_3$OH maser detection (M1) in Mol 121 is associated with MHO 864 (Figure 5). It is by far the brightest at 95 GHz and was the only previously detected maser at 44 GHz (S. Kurtz, private communication). MHO 865 and the associated maser (M3) are positioned exactly opposite core D, while MHO 892 and its associated maser (M2) lie along a slightly different position angle, though they also appear to be associated with core D. The masers all have low velocities, close to the V$_{LSR}$ of the ambient cloud, which is reported to be 1.5 km s$^{-1}$ by Zhang et al. (2005) and listed as 2.1 km s$^{-1}$ in the RMS catalog  (Mottram et al. 2011). This is consistent with the suggestion that CH$_3$OH masers arise in systemic gas in outflow cavity walls (DeBuizer et al. 2009).  

WSS12 suggested that the 95-GHz/44-GHz ratio might prove to be a useful diagnostic for the shortest-lived (e.g., very early) phases of massive star formation. In order to estimate the flux ratios of our masers, we binned the 44-GHz data (provided by S. Kurtz) to approximate our wider velocity channels. This summed data lowered the noise and revealed 44-GHz maser emission at the location of all three of our masers, not just the one known previously. We were then able to measure and compare the fluxes at 44 GHz and 95 GHz. These fluxes are all presented in Table 5. All three masers are brighter at 95 GHz than at 44 GHz, in contrast to the statistical result reported by Fontani et al. (2010), who concluded that the 95-GHz line is intrinsically fainter based on their detection rates. M1 and M3, which are associated with the same outflow angle, have remarkably similar flux ratios (3.73 and 3.69, respectively). M2, which is associated with MHO 892,  has a 95-GHz to 44-GHz flux ratio of 6.63, dramatically higher than the other two. MHO 892 is much more compact than MHOs 864 \& 865, though we do not know if this is related in any way.

 \subsection{Different Stages of Massive Star Formation in Mol 121}

We can tentatively order the dense cores in Mol 121 by evolutionary stage, based on their continuum and spectral-line properties. (1) Core A is associated with the {\it IRAS} source, a compact PDR, and bright cm-continuum emission, suggesting it is the oldest site of massive star formation in Mol 121. (2) Core D, with its associated faint cm-continuum emission, luminous H$_2$ outflow, CH$_3$OH masers, and \ion{Fe}{2} emission may harbor one or more massive objects that has not yet destroyed its accretion disk. (3) Core C, with its DES, likely contains a MYSO in a pre-UC \ion{H}{2} region phase of evolution. (4) B \& E appear to be massive starless or prestellar cores. We note that cores B \& C also lie near the periphery of the PDR, which suggests that Mol 121 would be an excellent candidate for follow-up studies to explore the possibility of star formation triggered by the expanding \ion{H}{2} region associated with {\it IRAS} 20188$+$3928. 

\section
{SUMMARY AND CONCLUSIONS}

\begin{enumerate}

\item Using NICFPS on the ARC 3.5-m telescope at APO, we have discovered 12 new MHOs in the vicinity of Mol 121/{\it IRAS} 20188$+$3928, bringing the total known MHOs in this region to 17. We have been able to separate one of the five previously discovered MHOs from a PDR and a region marked by an anomalous H$_2$ 2.12-$\micron$/2.25-$\micron$ flux ratio $<$ 1. To the best of our knowledge, this is the first time an observation of a H$_2$ 2.12-$\micron$/2.25-$\micron$ flux ratio $<$ 1 has been reported.

\item We have detected  [\ion{Fe}{2}] 1.64-$\micron$ emission toward MHOs 864 \& 867. [\ion{Fe}{2}] 1.64-$\micron$/H$_2$ 2.12-$\micron$ and H$_2$ 2.12-$\micron$/2.25-$\micron$ flux ratios measured along and across these MHOs indicate the presence of both J-shock and C-shock components. 

\item CARMA 3-mm continuum and line observations have revealed the presence of five compact cores (A, B, C, D, \& E) and three CH$_3$OH masers (M1, M2, \& M3) in the Mol 121/{\it IRAS} 20188$+$3928 region. The physical properties of the cores indicate that  cores A, C, \& D are presently forming massive stars, while cores B \& E may be the sites of future massive star formation. Masers M1, M2, \& M3 lie toward MHOs 864, 892, \& 865; all of which are associated with outflow activity from one or more sources embedded in core D. The 95-GHz masers all have 44-GHz counterparts, but they are all brighter at 95 GHz. The two masers (M1 \& M3) associated with bipolar MHOs 864 \& 865 have strikingly similar ratios of $\sim$ 3.7, while M2 has a higher ratio of $\sim$ 6.6. The significance of this is presently unknown, but suggests that future studies of maser line ratios are warranted, and may hold important implications for probing physical conditions in massive star-forming regions.

\item The positions and orientations of the MHOs relative to the cores suggest the presence of multiple outflows originating from at least two of the cores. A comparison of the total H$_2$ luminosity of the MHOs to the total bolometric luminosity of this region strongly suggests the outflows are driven by massive YSOs.

\item We tentatively age-order the cores from most to least evolved based on the different signposts of star-formation activity. 
(1) Core A is associated with the {\it IRAS} source, a compact PDR, and bright cm-continuum emission, suggesting it is the oldest site of massive star formation in Mol 121. (2) Core D, with its associated faint cm-continuum emission, luminous well-collimated H$_2$ outflow(s), CH$_3$OH masers, and \ion{Fe}{2} emission may harbor one or more massive objects that has not yet destroyed its accretion disk. (3) Core C is not associated with cm-continuum emission, but it coincides with a DES, the anomalous H$_2$ 2.12-$\micron$/2.25-$\micron$ emission, and possible outflow. It likely harbors a MYSO in a pre-UC \ion{H}{2} region phase of evolution. (4) While cores B \& E are massive, they also appear to be starless or prestellar. 

 \end{enumerate}

\acknowledgments

This research is based on observations obtained with the Apache Point Observatory 3.5-meter telescope, which is owned and operated by the Astrophysical Research Consortium. We particularly thank the 3.5-meter Observing Specialists for their assistance in acquiring the data; Al Harper, RS's advisor during acquisition of the CARMA observations; and Michael Medford, who performed the original NICFPS data reduction for Mol 121. Support for CARMA construction was derived from the Gordon and Betty Moore Foundation, the Kenneth T. and Eileen L. Norris Foundation, the James S. McDonnell Foundation, the Associates of the California Institute of Technology, the University of Chicago, the states of California, Illinois, and Maryland, and the National Science Foundation. Ongoing CARMA development and operations are supported by the National Science Foundation under a cooperative agreement, and by the CARMA partner universities. Special thanks go to Stan Kurtz, for supplying us with his VLA CH$_3$OH 44 GHz maser data and to Michael Smith, for useful discussions on shock physics and the possible origin of the anomalous H$_2$ emission reported in this paper.  This research has made use of SAOImage DS9, developed by the Smithsonian Astrophysical Observatory. We acknowledge use of the NASA/IPAC Infrared Science Archive, which is operated by the Jet Propulsion Laboratory, California Institute of Technology, under contract with the National Aeronautics and Space Administration. This publication makes use of  data products from the Two Micron All Sky Survey, which is a joint project of the University of Massachusetts and the Infrared Processing and Analysis Center/California Institute of Technology, funded by the National Aeronautics and Space Administration and the National Science Foundation; and the Wide-field Infrared Survey Explorer, which is a joint project of the University of California, Los Angeles, and the Jet Propulsion Laboratory/California Institute of Technology, funded by the National Aeronautics and Space Administration. GW-C was funded in part through NASA's Illinois Space Grant Consortium, and the authors gratefully acknowledge support from the Brinson Foundation grant in aid of astrophysics research at the Adler Planetarium. KA thanks Jamie Riggs for help in developing the near-infrared continuum subtraction procedure. We also thank the anonymous referee for providing helpful and timely comments to improve this paper.

\clearpage
\begin{figure}
\plotone{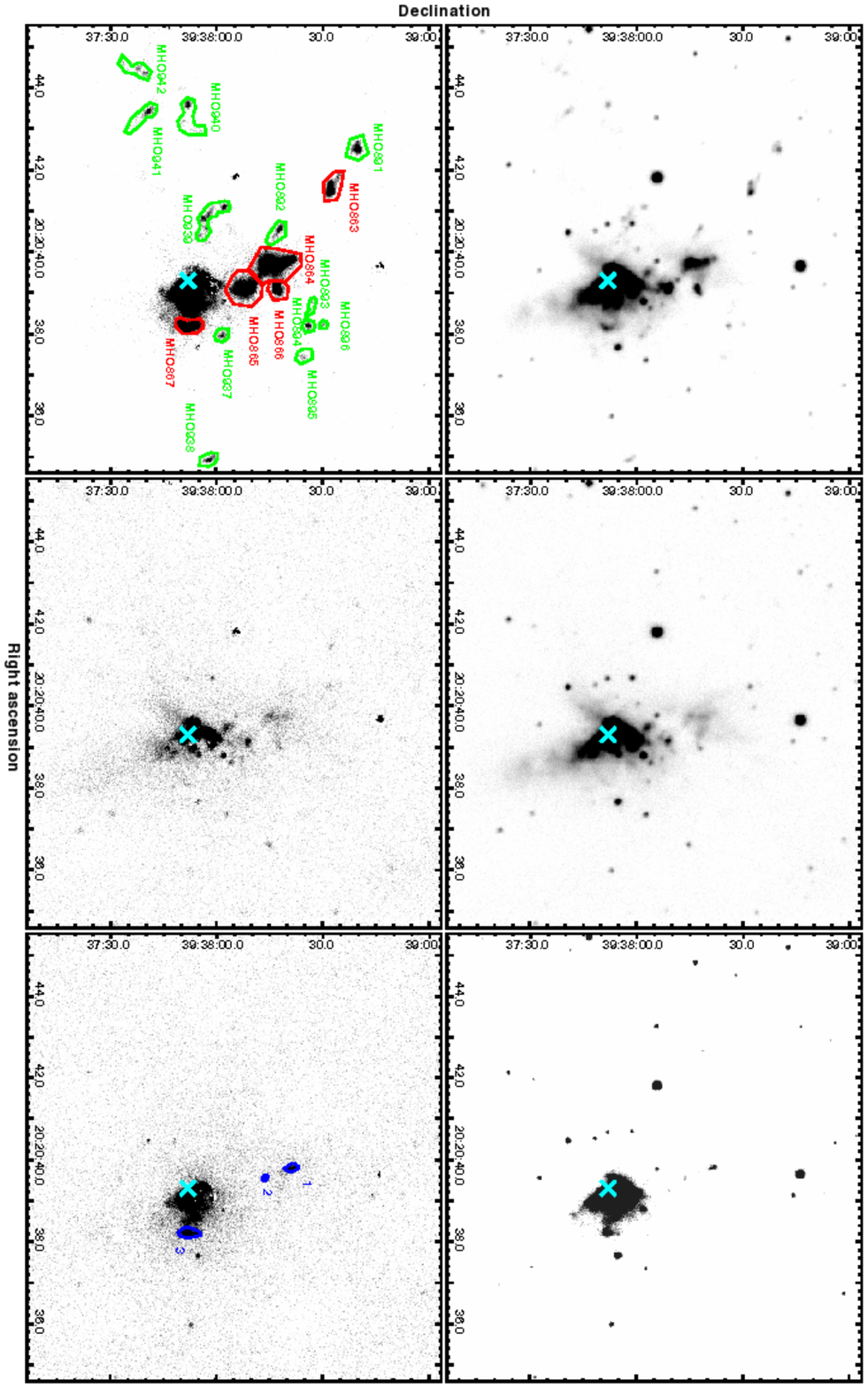}
\end{figure}

\clearpage
\begin{figure}
\plotone{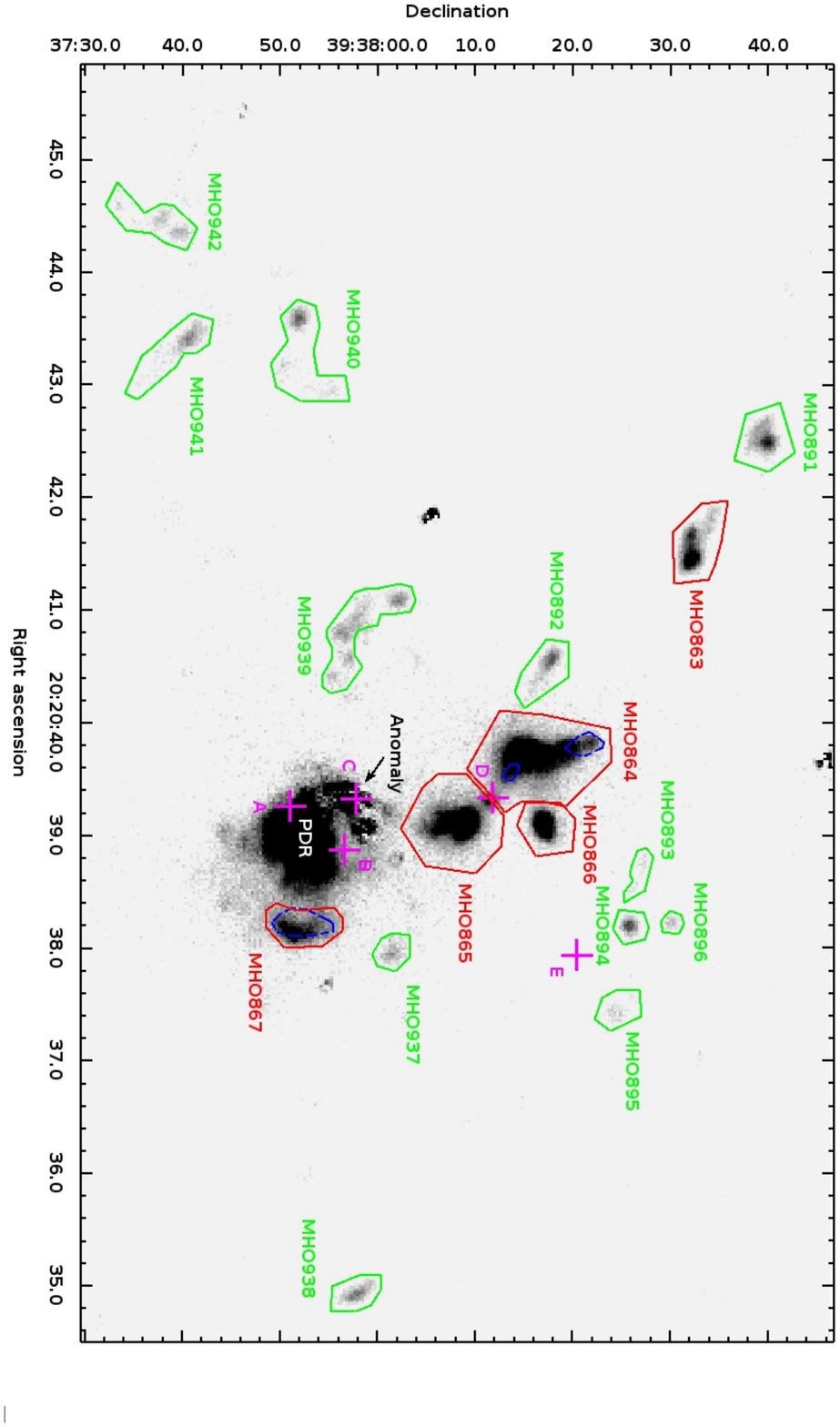}
\end{figure}

\clearpage
\begin{figure}
\plotone{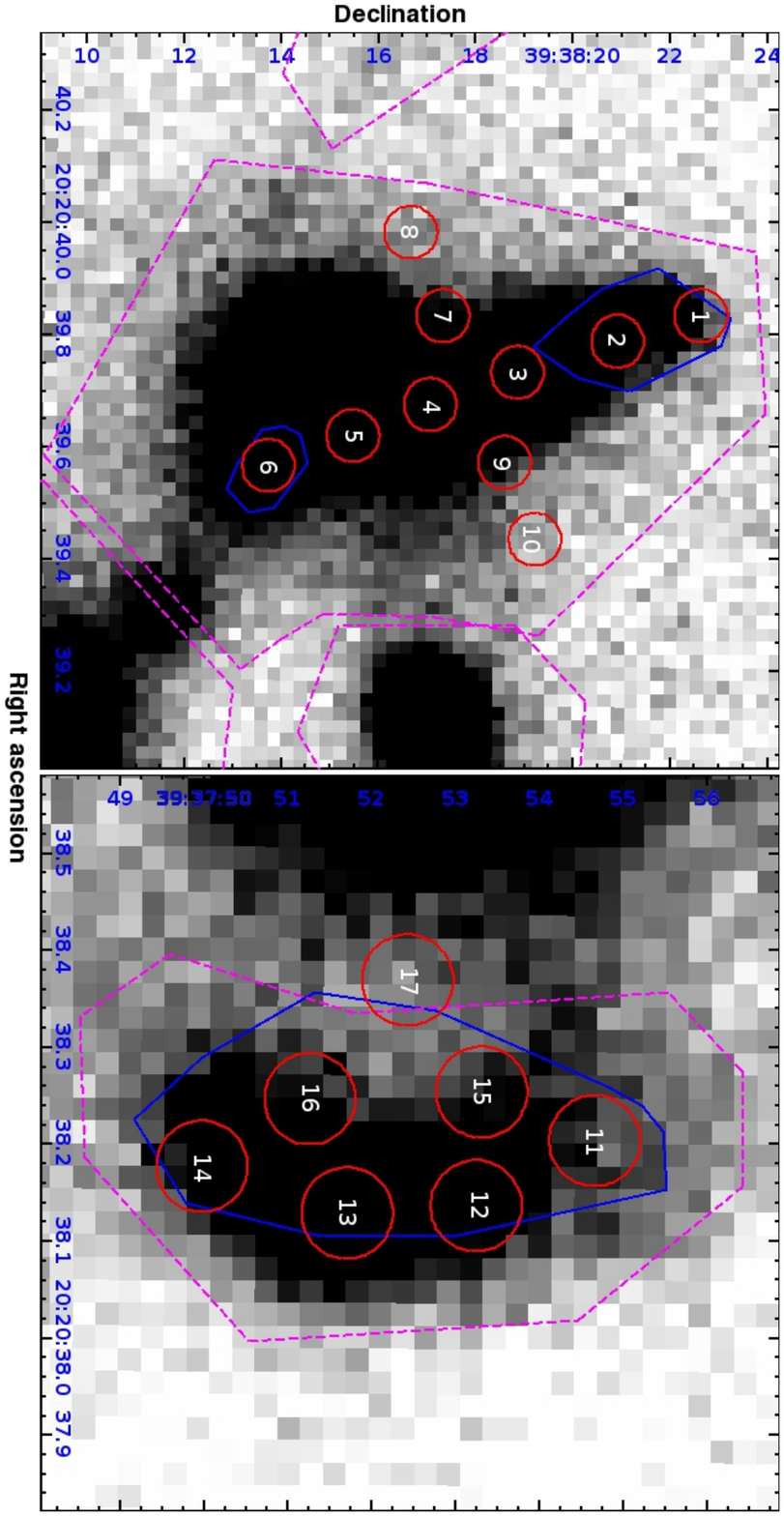}
\end{figure}

\clearpage
\begin{figure}
\plotone{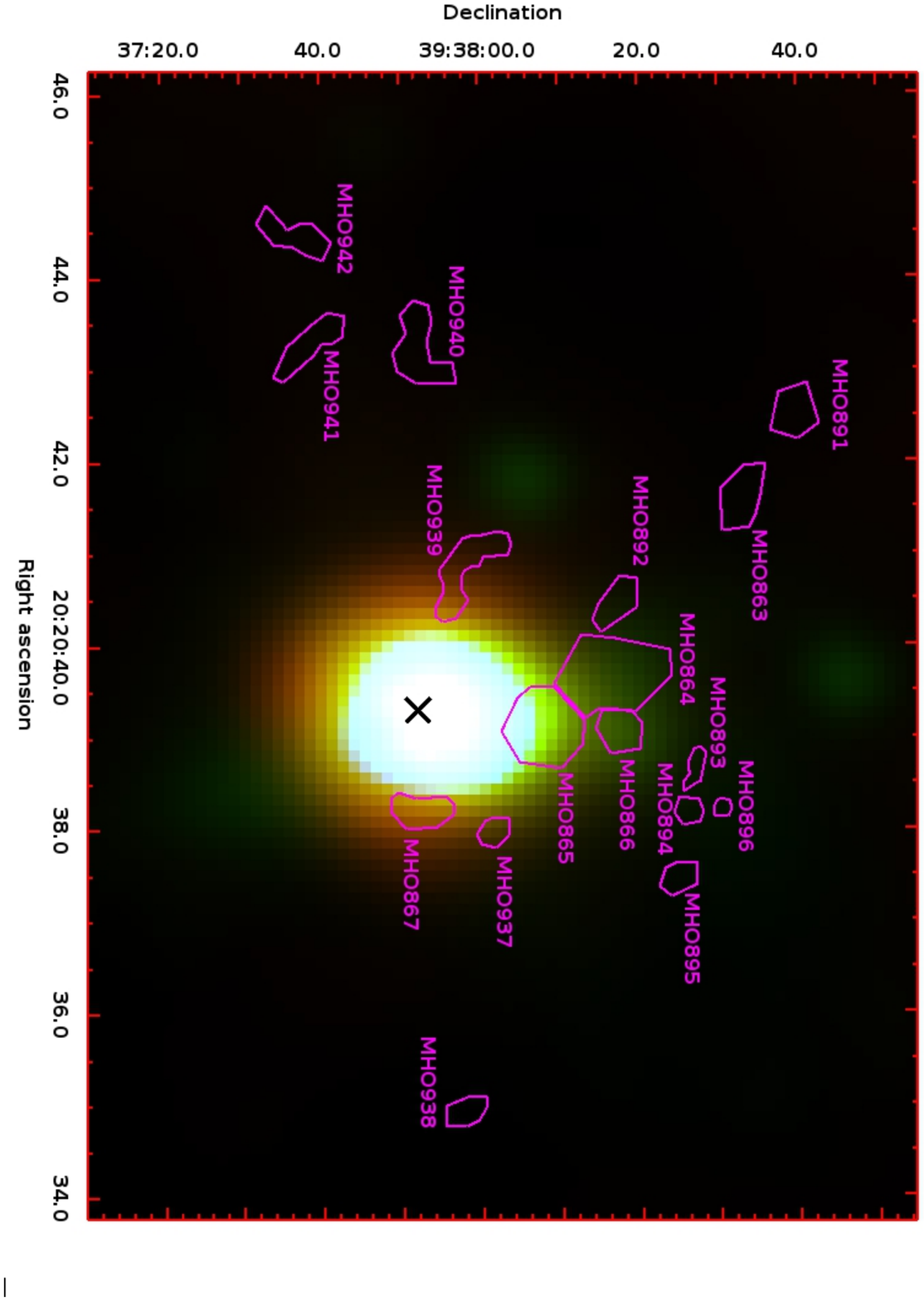}
\end{figure}

\clearpage
\begin{figure}
\plotone{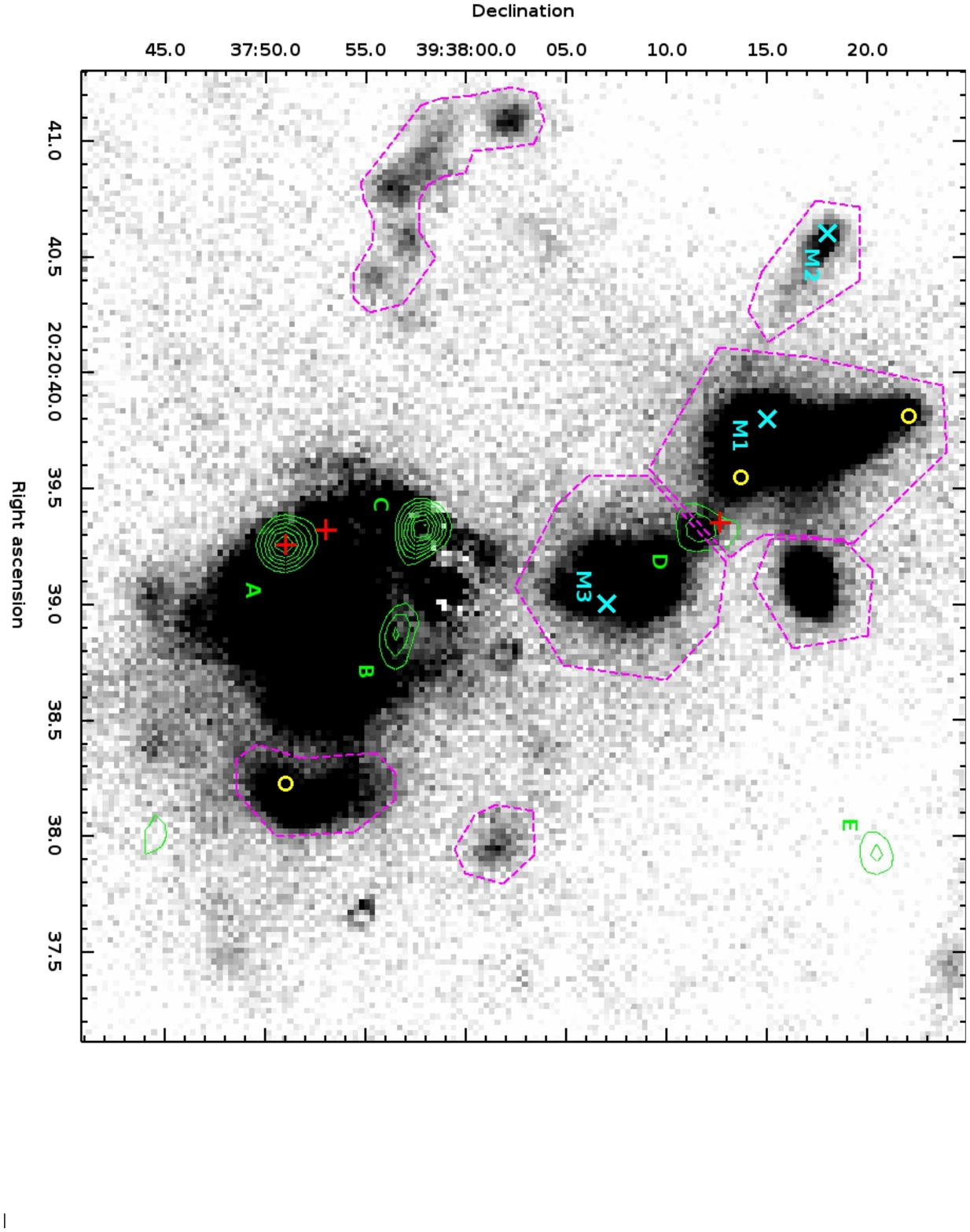}
\end{figure}

\clearpage

\begin{deluxetable}{rrcccccccc}
\rotate
\footnotesize
\tablecaption{Mol 121 H$_2$ Emission Line Fluxes}
\label{table1}
\tablewidth{0pt}
\tabletypesize{\scriptsize}
\tablehead{
\colhead{MHO \#} &  \colhead{Area} & \colhead{$\alpha$(J2000)\tablenotemark{a}}  & \colhead{$\delta$(J2000)\tablenotemark{a}} &  \colhead{F$_{2.12 \mu m}$\tablenotemark{b}} &  \colhead{F$_{2.25 \mu m}$\tablenotemark{b}} & \colhead{F$_{2.12}$/F$_{2.25}$\tablenotemark{c}}  & \colhead{F$_{2.12 \mu m}$(peak)\tablenotemark{b,d}}  &  \colhead{F$_{2.25 \mu m}$(peak)\tablenotemark{b,d}} &  \colhead{F$_{2.12}$/F$_{2.25}$(peak)\tablenotemark{c}}  \\
 \colhead{}   &   \colhead{$10^{-10}$~Sterad}   &  \colhead{(h m s)}  &  \colhead{($^{\circ}$ $^{\prime}$ $^{\prime\prime}$)}  &  \colhead{10$^{-18}$ W m$^{-2}$}  & 
      \colhead{10$^{-18}$ W m$^{-2}$} &  \colhead{}  &  \colhead{10$^{-18}$ W m$^{-2}$} &  \colhead{10$^{-18}$ W m$^{-2}$}  &  \colhead{}  \\
      }
\startdata
863\tablenotemark{e}  &  8.944 & 20 20 41.5  &  +39 38 32 &  27.4$\pm$2.1 &  1.3$\pm$0.3  &  20.4$\pm$4.3  &  4.49$\pm$0.34 &   0.40$\pm$0.05 &  11.2$\pm$1.5 \\
864\tablenotemark{e}  &  26.627 & 20 20 39.6 &  +39 38 16 &  158.3$\pm$13.1 &  17.4$\pm$2.7  &  9.1$\pm$1.6 &  10.20$\pm$0.77 &  0.92$\pm$0.10  & 11.0$\pm$1.4   \\
865\tablenotemark{e}  &  19.225 & 20 20 39.0 & +39 38 09  &  99.4$\pm$8.5 &  19.2$\pm$2.1  &  5.2$\pm$0.7 &  9.28$\pm$0.70 &   2.09$\pm$0.16  &   4.4$\pm$0.5  \\
866\tablenotemark{e}  &  6.648 &  20 20 39.0  & +39 38 16 &  33.0$\pm$2.7 &  2.6$\pm$0.5  &  12.7$\pm$2.7  & 2.23$\pm$0.20  &   0.46$\pm$0.08  &  4.9$\pm$1.0  \\
867\tablenotemark{e,f}  &  7.179 & 20 20 38.1 &  +39 37 52  &   42.2$\pm$3.2 &  2.1$\pm$0.4  &  20.1$\pm$4.5 &  3.03$\pm$0.27 &  0.33$\pm$0.10  &   9.1$\pm$2.9  \\
891    &  7.676 &  20 20 42.5 & +39 38 40 &  14.9$\pm$1.1 &  0.7$\pm$0.2  &  21.0$\pm$6.6  &  2.59$\pm$0.20 &  0.18$\pm$0.04 &  14.6$\pm$3.3  \\
892    &  6.186 &  20 20 40.5 & +39 38 18  &   9.0$\pm$0.9 &  \nodata  &  $>$7.5  & 1.79$\pm$0.14 & \nodata  &  $>$ 10.1   \\
893    &  2.724 & 20 20 38.7  & +39 38 27 &    2.3$\pm$0.3 &  \nodata  &  $>$5.6 &  0.39$\pm$0.05  &  \nodata  &    $>$ 2.6  \\
894    &  2.707 & 20 20 38.2 & +39 38 26 &   5.4$\pm$0.4 &  \nodata  &  $>$16.7  & 2.01$\pm$0.15 &   \nodata  &  $>$ 14.0  \\
895     &  3.941 & 20 20 37.5  & +39 38 25 &   3.2$\pm$0.4 &  \nodata  &  $>$5.4 & 0.51$\pm$0.06  &  \nodata  &   $>$ 3.5  \\
896    &  1.268  & 20 20 38.2  & +39 38 30  &  1.7$\pm$0.2 &  0.3$\pm$0.1  &  5.4$\pm$1.4  & 0.68$\pm$0.06  &  0.18$\pm$0.05 &    3.9$\pm$1.2  \\
937    &  3.187 &  20 20 37.9  &  + 39 38 02 &  4.2$\pm$0.6 &  \nodata  &  $>$7.0 & 0.70$\pm$0.12  &  \nodata  &   $>$ 2.4   \\
938    &  4.078  &20 20 34.9  & +39 37 58  &  6.8$\pm$0.5 &  \nodata  &  $>$16.1  &  1.61$\pm$0.13  &  0.13$\pm$0.04  &   12.7$\pm$3.8  \\
939     & 11.360  & 20 20 41.1  & +39 38 02 &  19.0$\pm$2.8 &  \nodata  &  $>$2.0 & 1.13$\pm$0.11  &  \nodata  &    $>$ 7.3  \\
940    &  11.669  & 20 20 43.6  & +39 37 52 &   13.1$\pm$1.0 &  \nodata  &  $>$15.2  & 1.75$\pm$0.14  &  \nodata  &   $>$ 14.7  \\
941     &  8.430 & 20 20 43.4  &  +39 37 41 & 11.1$\pm$0.9 &  0.9$\pm$0.2  &  12.0$\pm$3.0  & 1.34$\pm$0.11  &   0.12$\pm$0.04 &  11.6$\pm$4.3  \\
942    &  7.642 &  20 20 44.4  &  +39 37 40 &  9.5$\pm$0.7 &  \nodata  &  $>$16.4  &  0.60$\pm$0.07 &  \nodata  &   $>$ 5.0  \\
Anomaly\tablenotemark{g}  &  28.718 &  \nodata & \nodata  & 124.9$\pm$8.5 &  269.0$\pm$16.6  &  0.46$\pm$0.04   &  \nodata   & \nodata &   \nodata \\
PDR\tablenotemark{g}   &  95.886 &  \nodata & \nodata  & 552.5$\pm$37.8 &  356.9$\pm$23.7  &  1.5$\pm$0.1 &  \nodata   &\nodata  &  \nodata     \\
\enddata
\tablenotetext{a}{Coordinates for MHOs 863-867 are taken from the on-line catalog (Davis et al. 2010); coordinates for the MHOs discovered in this research correspond to H$_2$ 2.12 $\micron$ emission peaks.}
\tablenotetext{b}{Flux errors combine three categories of uncertainty (calibration, subtraction, and aperture) and  correspond to $\pm$1 $\sigma$ (see \S 2.1).}
\tablenotetext{c}{See \S 3.1 for procedure used to determine F$_{2.25\micron}$ upper limits.}
\tablenotetext{d}{Peaks are calculated using a circular aperture of radius 2 pixels centered on maximum 2.25 $\micron$ emission ($\sim 0.3 \times 10^{-10}$ sterad).} 
\tablenotetext{e}{MHOs first observed by Varricatt et al. (2010). } 
\tablenotetext{f}{Originally, the feature identified by Varricatt et al. (2010) as MHO 867 encompassed the PDR as well as the bow-shaped MHO we identify in this paper. This entry redefines the position and extent of MHO 867 to include the shocked region only.} 
\tablenotetext{g}{These are not shock features, and therefore not MHOs.}

\end{deluxetable}

\clearpage

\begin{deluxetable}{ccccc}
\footnotesize
\tablecaption{Mol 121 [\ion{Fe}{2}]  Emission Line Fluxes}
\label{table1}
\tablewidth{0pt}
\tabletypesize{\scriptsize}
\tablehead{
 \colhead{Associated MHO} &  \colhead{Area} & 
\colhead{$\alpha$(J2000)\tablenotemark{a}}  & \colhead{$\delta$(J2000)\tablenotemark{a}} &
\colhead{F$_{1.64 \mu m}$ \tablenotemark{b}}  \\
 \colhead{}  &   \colhead{$10^{-10}$~Sterad} &   \colhead{(h m s)}  &  \colhead{($^{\circ}$ $^{\prime}$ $^{\prime\prime}$)}   &  10$^{-18}$ W m$^{-2}$  \\
      }
\startdata
864  &  1.833 &  20 20 39.8  & +39 38 22  &  7.04$\pm$1.17   \\
  &  0.668 & 20 20 39.5 & +39 38 14 &   2.14$\pm$0.36   \\
867 &  3.564 & 20 20 38.2 & +39 37 51 &  19.89$\pm$4.07   \\
\enddata
\tablenotetext{a}{Coordinates correspond to emission peaks.}
\tablenotetext{b}{Flux errors combine three categories of uncertainty (calibration, subtraction, and aperture) and  correspond to $\pm$1 $\sigma$ (see \S 2.1).}
\end{deluxetable}

\clearpage

\begin{deluxetable}{cccccc}
\footnotesize
\tablecaption{MHO 864 \& MHO 867 Emission Line Flux Ratios}
\label{table1}
\tablewidth{0pt}
\tabletypesize{\scriptsize}
\tablehead{
 \colhead{MHO \#} &  \colhead{Aperture \#} & 
\colhead{$\alpha$(J2000)\tablenotemark{a}}  & \colhead{$\delta$(J2000)\tablenotemark{a}} &
\colhead{F$_{1.64}$/F$_{2.12}$}    &  \colhead{F$_{2.12}$/F$_{2.25}$} \\
 \colhead{}  &   \colhead{} &   \colhead{(h m s)}  &  \colhead{($^{\circ}$ $^{\prime}$ $^{\prime\prime}$)}   &    &     \\
      }
\startdata
864  &  1  &  20 20 39.8  &  +39 38 23  &  0.9$\pm$0.2 &  $>$5.9   \\
  &  2 & 20 20 39.8 & +39 38 21 & 0.8$\pm$0.1  & 7.0$\pm$1.8  \\
   &  3 & 20 20 39.7 & +39 38 19 & 0.1$\pm$0.03  & 6.8$\pm$1.3 \\
    &  4 & 20 20 39.7 & +39 38 17 & $<$0.03   & 13.8$\pm$2.8  \\
     &  5 & 20 20 39.6 & +39 38 15 & $<$0.03  & 14.5$\pm$4.5 \\
      &  6 & 20 20 39.6 & +39 38 14 & 0.4$\pm$0.1  & 12.8$\pm$4.3  \\
       &  7 & 20 20 39.8 & +39 38 17 & $<$0.04  &  $>$7.9 \\
 &  8 & 20 20 40.0 & +39 38 17 & $<$0.3  & $>$1.4  \\
  &  9 & 20 20 39.6 & +39 38 19 &  $<$0.03 &  $>$7.1  \\
   &  10 & 20 20 39.4 & +39 38 19 & \nodata\tablenotemark{b}  & $>$0.7 \\
867 &  11 & 20 20 38.2 & +39 37 55 & 1.2$\pm$0.3 & $>$4.0  \\
 &  12 & 20 20 38.1 & +39 37 53 &  0.5$\pm$0.1 & $>$7.2 \\
  &  13 & 20 20 38.1 & +39 37 52 &  0.6$\pm$0.1 & 10.4$\pm$3.6 \\
   &  14 & 20 20 38.2 & +39 37 50 & 0.5$\pm$0.1  &  $>$6.5  \\
    &  15 & 20 20 38.3 & +39 37 53 &  1.2$\pm$0.2 & $>$4.5  \\
     &  16 &  20 20 38.2 & +39 37 51 & 1.8$\pm$0.3  &  $>$5.6 \\
      &  17 & 20 20 38.4 & +39 37 52 &  1.9$\pm$0.5 &  $>$2.3 \\
\enddata
\tablenotetext{a}{Coordinates of aperture centers.}
\tablenotetext{b}{$\sigma <3$}
\end{deluxetable}

\clearpage

\begin{deluxetable}{ccccccccccc}
\tabletypesize{\scriptsize}
\rotate
\footnotesize
\tablecaption{Properties of 3-mm Cores}
\label{table1}
\tablewidth{0pt}
\tablehead{
Core & $\alpha$(J2000) & $\delta$(J2000) & Size\tablenotemark{a} &  I$_{93.6 GHz}$ &
S$_{93.6 GHz}$  & M$_{core}$(T$_d$=20 K)  & M$_{core}$(T$_d$=40 K)   &  N$_{H_2}$(T$_d$=20 K)  &  N$_{H_2}$(T$_d$=40 K)  \\
      &   (h m s)  & ($^{\circ}$ $^{\prime}$  $^{\prime\prime}$) & ($^{\prime\prime} \times ^{\prime\prime}$) &  (mJy beam$^{-1}$) & (mJy) & ( M$_{\odot}$)  & ( M$_{\odot}$)    & ($\times 10^{24}$ cm$^{-2}$) & ($\times 10^{24}$ cm$^{-2}$)   \\
      }
\startdata
A\tablenotemark{b} &  20 20 39.262  &  +39 37 51.018 &  1.36$\times$0.31  & 13.6 & 17.4 & 29.9 & 14.1 &  6.00 &  2.83  \\
B &  20 20 38.865  & +39 37 56.635 &  2.60$\times$0.08 & 7.3 & 12.0  & 20.6 & 9.74 &  3.22 &  1.52  \\
C\tablenotemark{c} & 20 20 39.325  & +39 37 57.770  &  1.16$\times$0.45  &  15.0 & 18.7  &  32.1  &  15.2  & 6.62  &  3.12  \\
D\tablenotemark{d} & 20 20 39.330   &  +39 38 11.770 &  2.87$\times$1.50  & 7.3 & 17.4 & 29.9  & 14.1   & 3.22  &  1.52  \\
E &  20 20 37.935  &  +39 38 20.393  &  2.83$\times$0.61 &  6.4 & 10.5 & 18.1  & 8.55   &  2.82 & 1.33  \\
\enddata
\tablenotetext{a}{Size after de-convolution from the CARMA beam.}
\tablenotetext{b}{Coincident with the {\it IRAS} source, 24.2 mJy \& 2.2 mJy 3.6-cm continuum emission detected by Jenness et al. (1995), 29.95 mJy 6-cm continuum emission detected by Molinari et al. (1998), and H$_2$O maser emission (Palla et al. 1991; Brand et al. 1994; Anglada et al. 1997).}
\tablenotetext{c}{Coincident with a DES reported by Yao et al. (2000) and anomalous H$_2$ emission discussed in the text.}
\tablenotetext{d}{Coincident with a H$_2$O maser and 3.4 mJy 3.6-cm continuum emission detected by Jenness et al. (1995), and 3.82 mJy 6-cm continuum emission detected by Molinari et al. (1998).}

\end{deluxetable}

\clearpage

\begin{deluxetable}{cccccccc}
\tabletypesize{\scriptsize}
\rotate
\footnotesize
\tablecaption{CH$_3$OH Maser Properties}
\label{table 2}
\tablewidth{0pt}
\tablehead{
Maser   & Associated MHO  & $\alpha$(J2000)   & $\delta$(J2000) 
 &  $S_{95} \Delta V$  &  V$_{LSR}$\tablenotemark{a}   &   $S_{44} \Delta V$\tablenotemark{b}    &    $S_{95} \Delta V$/$S_{44} \Delta V$
  \\
 &   &  (h m s)  & ($^{\circ}$ $^{\prime}$  $^{\prime\prime}$)  & (Jy km s$^{-1}$) &  (km s$^{-1}$)    &  (Jy km s$^{-1}$)   &                        \\
      }
\startdata
 M1  &  864  &
20 20 39.8   & 39 38 15 & 5.29
&  1.8  &  1.42     &  3.73  \\
 M2  &  892  &
20 20 40.6   & 39 38 18 & 1.79
&  0.5  & 0.27   & 6.63   \\
 M3  &  865  &
20 20 39.0  & 39 38 07 & 1.18
&  0.1  &  0.32   & 3.69  \\
\enddata
\tablenotetext{a}{Central V$_{LSR}$ of the 1.538 km s$^{-1}$ wide channel 
 95-GHz intensity peak.}
 \tablenotetext{b}{The 0.166 km s$^{-1}$ 44-GHz data were binned to the 1.538 km s$^{-1}$ velocity resolution of the 95-GHz data.}

\end{deluxetable}

\clearpage
\figcaption[Mol121_fig1.eps]{Inverted greyscale images of H$_2$ 2.12-$\mu$m (top left),  H$_2$ 2.25-$\mu$m (top middle), and   [\ion{Fe}{2}] 1.64-$\micron$ emission (top right);  and continuum-subtracted  H$_2$ 2.12-$\mu$m (bottom left),  H$_2$ 2.25-$\mu$m (bottom middle), and  [\ion{Fe}{2}] 1.64-$\micron$ (bottom right) emission in the Mol 121 region. The cyan X marks the position of the {\it IRAS} source. MHOs reported by Varricatt et al. (2010) are labelled and enclosed in red and MHOs reported in this paper are labelled and enclosed in green on the continuum-subtracted H$_2$ 2.12-$\mu$m image. [\ion{Fe}{2}] knots are enclosed with blue dashed lines on the continuum-subtracted [\ion{Fe}{2}] 1.64-$\micron$ image.
}

\figcaption[Mol121_fig2.eps]{Inverted greyscale image of  continuum-subtracted H$_2$ 2.12-$\mu$m emission.  MHOs reported by Varricatt et al. (2010) are labelled and enclosed in red, MHOs reported in this paper are labelled and enclosed in green, [\ion{Fe}{2}] knots are enclosed with blue dashed lines, and peaks of 3-mm cores identified with CARMA are indicated with magenta crosses and labelled as discussed in the text. The PDR and anomalous H$_2$ emission (Anomaly) are labelled.
}

\figcaption[Mol121_fig3.eps]{Inverted greyscale image of  continuum-subtracted H$_2$ 2.12-$\mu$m emission in the vicinity of  MHO 864 (left) and MHO 867 (right).  MHOs and [\ion{Fe}{2}] knots are enclosed as in Figure 2. Aperture regions corresponding to entries in Table 3 are indicated and labelled.
}

\figcaption[Mol121_fig4.eps]{Three-color WISE image of the Mol 121 region. Colors are 3.4 $\micron$ (blue), 4.6 $\micron$ (green), and 12 $\micron$ (red). MHOs are indicated. The black X denotes the position of the {\it IRAS} source.
}

\figcaption[Mol121_fig5.eps]{Inverted greyscale of  continuum-subtracted H$_2$ 2.12-$\mu$m sub-image. CARMA 3-mm continuum cores (labelled green contours), [\ion{Fe}{2}] emission peaks (yellow circles), and 95-GHz CH$_3$OH masers (labelled cyan Xs) are indicated. Red crosses show the positions of VLA 3.6-cm sources reported by Jenness et al. (1995).
}


\begin{thebibliography}{}

\bibitem[\protect\citeauthoryear
{Allen}{2000}]{all00}
Allen, C. W. 2000, Allen's Astrophysical Quantities, 4th ed. (London: The Athlone Press, Ltd.),
ed. A. N. Cox, p. 527

\bibitem[\protect\citeauthoryear
{Anderson et al.}{2012}]{and12}
Anderson, L. D., Zavagno, A., Deharveng, L., et al. 2012, A\&A, 542, 10

\bibitem[\protect\citeauthoryear
{Anglada et al.}{1997}]{ang97}
Anglada, G., Sepulveda, I., \& Gomez, J. F. 1997, A\&AS, 121, 255

\bibitem[\protect\citeauthoryear
{APO}{2010}]{apol0}
Apache Point Observatory 2010, NICFPS (Sunspot, NM: APO),
 http://www.apo.nmsu.edu/arc35m/Instruments/NICFPS/

\bibitem[\protect\citeauthoryear
{Arvidsson et al.}{2010}]{arvl0}
Arvidsson, K., Kerton, C. R., Alexander, M. J., Kobulnicky, H. A., \& Uzpen, B. 2010,
AJ, 140, 462

\bibitem[\protect\citeauthoryear
{Barsony et al.}{2010}]{bar10}
Barsony, M., Wolf-Chase, G. A., Ciardi, D. R., \& O'Linger, J. 2010, ApJ, 720, 64

\bibitem[\protect\citeauthoryear
{Benjamin et al.}{2003}]{ben03}
Benjamin, R. A., Churchwell, E., Babler, B. L., et al. 2003, PASP, 115, 953

\bibitem[\protect\citeauthoryear
{Black \& van Dishoeck}{1987}]{bla87}
Black, J. H. \& van Dishoeck, E. F. 1987, IAU Symposium 115, 139

\bibitem[\protect\citeauthoryear
{Brand et al.}{1947}]{bra94}
Brand, J., Cesaroni, R., Caselli, P., et al. 1994, A\&AS, 103, 541

\bibitem[\protect\citeauthoryear
{Caratti o Garatti et al.}{2008}]{car08}
Caratti o Garatti, A., Froebrich, D., Eisl{\"o}ffel, J., Giannini, T., \& Nisini, B. 2008, A\&A, 485, 137

\bibitem[\protect\citeauthoryear
{Caratti o Garatti et al.}{2006}]{car06}
Caratti o Garatti, A., Giannini, T., Nisini, B. \& Lorenzetti, D. 2006, A\&A, 449, 1077

\bibitem[\protect\citeauthoryear
{Chen et al.}{2012}]{che12}
Chen, C-C., Williams, J. P., \& Pandian, J. D. 2012, ApJ, 752, 102

\bibitem[\protect\citeauthoryear
{Churchwell et al.}{2009}]{chu09}
Churchwell, E., Babler, B. L., Meade, M. R., et al. 2009, PASP, 121, 213

\bibitem[\protect\citeauthoryear
{Curran \& Chrysostomou}{2007}]{cur07}
Curran, R. L., \& Chrysostomou, A. 2007, MNRAS, 382, 699

\bibitem[\protect\citeauthoryear
{Cyganowski et al.}{2009}]{cyg09}
Cyganowski, C. J., Brogan, C. L., Hunter, T. R., \& Churchwell, E. 2009, ApJ, 702, 1615

\bibitem[\protect\citeauthoryear
{Cyganowski et al.}{2008}]{cyg08}
Cyganowski, C. J., Whitney, B. A., Holden, E., et al. 2008, AJ, 136, 2391

\bibitem[\protect\citeauthoryear
{Davis et al.}{2010}]{dav10}
Davis, C. J., Gell, R., Khanzadyan, T., Smith, M. D., \& Jenness, T. 2010,
A\&A, 511, A24

\bibitem[\protect\citeauthoryear
{De Buizer et al.}{2009}]{deb09}
De Buizer, J. M., Redman, R. O., Longmore, S. N., Caswell, J., \& Feldman, P. A. 2009,
A\&A, 493, 127

\bibitem[\protect\citeauthoryear
{Dove et al.}{1987}]{dov87}
Dove, J. E., Rusk, A. C. M., Cribb, P. H., \& Martin, P. G. 1987, ApJ, 318, 379

 \bibitem[\protect\citeauthoryear
{Enoch et al.}{2007}]{eno07}
Enoch, M. L., Glenn, J., Evans, N. J. II, et al. 2007, ApJ, 666, 982

 \bibitem[\protect\citeauthoryear
{Fazio et al.}{2004}]{faz04}
Fazio, G. G., Hora, J. L., Allen, L. E., et al. 2004, ApJS, 154, 10

 \bibitem[\protect\citeauthoryear
{Fontani et al.}{2010}]{fon10}
Fontani, F., Cesaroni, R., \& Furuya, R. S. 2010, A\&A, 517, A56

\bibitem[\protect\citeauthoryear
{Gredel \& Dalgarno}{1995}]{gre95}
Gredel, R., \& Dalgarno, A. 1995, ApJ, 446, 852

 \bibitem[\protect\citeauthoryear
{Hollenbach \& McKee}{1989}]{hol89}
Hollenbach, D., \& McKee, C. F. 1989, ApJ, 342, 306

\bibitem[\protect\citeauthoryear
{Jenness et al.}{1995}]{jen95}
Jenness, T., Scott, P. F., \& Padman, R. 1995, MNRAS, 276, 1024

\bibitem[\protect\citeauthoryear
{Krumholz et al.}{2010}]{kru10}
Krumholz, M. R., Cunningham, A. J., Klein, R. I., \& McKee, C. F. 2010, ApJ, 713, 1120

\bibitem[\protect\citeauthoryear
{Krumholz \& McKee}{2008}]{kru08}
Krumholz, M. R., \& McKee, C. F. 2008, Nature, 451, 1082

\bibitem[\protect\citeauthoryear
{Lee et al.}{2012}]{lee12} 
Lee, H-T, Takami, M., Duan, H-Y, et al. 2012, ApJS, 200, 2

\bibitem[\protect\citeauthoryear
{Lee et al.}{2011}]{lee11} 
Lee, K. I., Looney, L. W., Klein, R., \& Wong, S. 2011, MNRAS, 415, 2790

\bibitem[\protect\citeauthoryear
{Little et al.}{1988}]{lit88} 
Little, L. T., Bergman, P., Cunningham, C. T., et al. 1988, A\&A, 205, 129

 \bibitem[\protect\citeauthoryear
{Lorenzetti et al.}{2002}]{lor02} 
Lorenzetti, D., Giannini, T., Vitali, F., Massi, F., \& Nisini, B. 2002, ApJ, 564, 839

\bibitem[\protect\citeauthoryear
{Matsakos et al.}{2009}]{mat09}
Matsakos, T., Massaglia, S., Trussoni, E., et al. 2009, A\&A, 502, 217

 \bibitem[\protect\citeauthoryear
{McCutcheon et al.}{1995}]{mcc95}
McCutcheon, W. H., Sato, T., Purton, C. R., Matthews, H. E., \& Dewdney, P. E. 1995, AJ, 110, 1762

\bibitem[\protect\citeauthoryear
{Molinari et al.}{1998}]{mol98}
Molinari, S., Brand, J., Cesaroni, R., Palla, F., \& Palumbo, G. G. C. 1998, A\&A, 336, 339

\bibitem[\protect\citeauthoryear
{Molinari et al.}{2008}]{mo08a}
Molinari, S., Faustini, F., Testi, L., et al. 2008, A\&A, 487, 1119  

 \bibitem[\protect\citeauthoryear
{Motte et al.}{ 2007}]{mot07}
Motte, F., Bontemps, S., Schilke, P., et al. 2007, A\&A, 476, 1243

 \bibitem[\protect\citeauthoryear
{Mottram et al.}{ 2011}]{mot11}
Mottram, J. C., Hoare, M. G., Urquhart, J. S., et al. 2011, A\&A, 525, A149

 \bibitem[\protect\citeauthoryear
{Ossenkopf \& Henning}{ 1994}]{oss94}
Ossenkopf, V. \& Henning, Th. 1994, A\&A, 291, 943

\bibitem[\protect\citeauthoryear
{Palla et al.}{1991}]{pal91}
Palla, F., Brand, J., Cesaroni, R., Comoretto, G., \& Felli, M. 1991, A\&A, 246, 249

\bibitem[\protect\citeauthoryear
{Pudritz et al.}{2006}]{pud06} 
Pudritz, R. E., Rogers, C. S., \& Ouyed, R. 2006, MNRAS, 365, 1131

\bibitem[\protect\citeauthoryear
{Rosolowsky et al.}{2010}]{ros10}
Rosolowsky, E., Dunham, M. K., Ginsburg, A., et al. 2010, ApJS, 188, 123

\bibitem[\protect\citeauthoryear
{Schnee et al.}{2009}]{sch09}
Schnee, S., \& Carpenter, J. M. 2009, ApJ, 698, 1456

\bibitem[\protect\citeauthoryear
{Shu et al.}{1994}]{shu94}
Shu, F., Najita, J., Ostriker, E., et al. 1994, ApJ, 429, 781

\bibitem[\protect\citeauthoryear
{Simon et al.}{2001}]{sim01}
Simon, R., Jackson, J. M., Clemens, D. P., Bania, T. M., \& Heyer, M. H. 2001, ApJ, 551, 747

\bibitem[\protect\citeauthoryear
{Smith}{1994}]{smi94}
Smith, M. D. 1994, A\&A, 289, 256

\bibitem[\protect\citeauthoryear
{Smith}{1995}]{smi95}
Smith, M. D. 1995, A\&A, 296, 789

\bibitem[\protect\citeauthoryear
{Smith et al.}{2003}]{smi03}
Smith, M. D., Khanzadyan, T., \& Davis, C. J. 2003, MNRAS, 339, 524

\bibitem[\protect\citeauthoryear
{Smith}{2012}]{smi12}
Smith, M. D. 2012, Astrophysical Jets and Beams (Cambridge: Cambridge University Press), pp. 116 - 117

\bibitem[\protect\citeauthoryear
{Turner et al.}{1977}]{tur77}
Turner, J., Kirby-Docken, K., \& Dalgarno, A. 1977, ApJS, 35, 281

\bibitem[\protect\citeauthoryear
{Varricatt et al.}{20010}]{var10}
Varricatt, W. P., Davis, C. J., Ramsay, S., \& Todd, S. P. 2010, MNRAS, 404, 661

 \bibitem[\protect\citeauthoryear
{Wolf-Chase et al.}{2012}]{wol12}
Wolf-Chase, G., Smutko, M., Sherman, R., Harper, D. A., \& Medford, M. 2012, ApJ, 745, 116 (WSS12)
 
\bibitem[\protect\citeauthoryear
{Wolf-Chase et al.}{2003}]{wol03}
Wolf-Chase, G., Moriarty-Schieven, G., Fich, M., \& Barsony, M. 2003, MNRAS, 344, 809

\bibitem[\protect\citeauthoryear
{Wolf-Chase et al.}{1995}]{wol95}
Wolf-Chase, G. A., Walker, C. K., \& Lada, C. J. 1995, ApJ, 442, 197

\bibitem[\protect\citeauthoryear
{Wood \& Churchwell}{1989}]{woo89} 
Wood, D. O. S., \& Churchwell, E. 1989, ApJ, 340, 265

\bibitem[\protect\citeauthoryear
{Wright et al.}{2010}]{wri10} 
Wright, E. L., Eisenhardt, P. R. M., Mainzer, A. K., et al. 2010, AJ, 140, 1868

\bibitem[\protect\citeauthoryear
{Yao et al.}{2000}]{yao00} 
Yao, Y., Ishii, M., Nagata, T., Nakaya, H., \& Sato, S. 2000, ApJ, 542, 392

\bibitem[\protect\citeauthoryear
{Zhang et al.}{2005}]{zha05}  
Zhang, Q., Hunter, T. R., Brand, J., et al. 2005, ApJ, 625, 864

\bibitem[\protect\citeauthoryear
{Zinnecker \& Yorke}{2007}]{zin07}
Zinnecker, H., \& Yorke, H. W. 2007, ARA\&A, 45, 481

\end{thebibliography}
\end{document}